\title{Testing our conceptual understanding of V1 function}
\author{
Urs K\"oster \\ 
Redwood Center for Theoretical Neuroscience \\
UC Berkeley 
\and 
Bruno A Olshausen \\ 
Redwood Center for Theoretical Neuroscience \\
UC Berkeley
}
 
\date{\today}

\documentclass[10pt]{article} % For LaTeX2e
\usepackage{times} % anonymized submission style files
\usepackage{amsmath,amssymb}
\usepackage{graphicx}
\usepackage{color}
\usepackage{sidecap}

\newcommand{\x}{ \mathbf{x}}

\newcommand{\mb}{\mathbf}

\begin{document}

\maketitle

\begin{abstract}

Here we test our conceptual understanding of V1 function by asking two experimental questions:  1) How do neurons respond to the spatiotemporal structure contained in dynamic, natural scenes?  and 2) What is the true range of visual responsiveness and predictability of neural responses obtained in an unbiased sample of neurons across all layers of cortex?  We address these questions by recording responses to natural movie stimuli with 32 channel silicon probes.  By simultaneously recording from cells in all layers, and taking all cells recorded, we reduce recording bias that results from ``hunting'' for neural responses evoked from drifting bars and gratings.  A nonparametric model reveals that many cells that are visually responsive do not appear to be captured by standard receptive field models. Using nonlinear Radial Basis Function kernels in a support vector machine, we can explain the responses of some of these cells better than standard linear and phase-invariant complex cell models. This suggests that V1 neurons exhibit more complex and diverse responses than standard models can capture, ranging from simple and complex cells strongly driven by their classical receptive fields, to cells with more nonlinear receptive fields inferred from the nonparametric and RFB model, and cells that are not visually responsive despite robust firing.

\end{abstract}

% ----------------------------------
% INTRODUCTION
% ----------------------------------
\section{Introduction}

Our current understanding of V1 function has been largely shaped by studying individual neurons with simple stimuli such as bars, spots, gratings or white noise.  Out of this body of work has emerged what might be called the standard model of V1:  an array of neurons essentially forming a bank of localized, oriented band-pass filters (with additional nonlinearities typically include some form of rectification and normalization).   This conceptual view has inspired a plethora models in computer vision\cite{Lowe1999, Serre2007}, and it forms the basis of the linear nonlinear Poisson (LNP) model~\cite{Simoncelli2004} which posits that the relevant output of a cell is the firing rate computed as a function of one or more linear filters applied to the stimulus.  Although it is often implied that a cell can be fully characterized this way, it is not at all obvious how well models from this family generalize to more ecologically valid stimuli, and to different cell types.  Our goal here is to explore the range of responses exhibited in an unbiased sample of V1 neurons recorded across all layers, in response to dynamic natural scenes.

The standard model depends on two assumptions:  Firstly, that neurons have ‘receptive fields’ that dictate how they respond to stimuli, and that these receptive fields may be mapped out with simple stimuli such as bars or gratings (or various spatial combinations thereof to account for nCRF effects); Secondly, that cortex is comprised of cells that share the same response properties, differing only in their specific tuning parameters.  We put these assumptions to the test in an exhaustive examination of V1 responses to natural movies. Using a recording technique that minimizes bias for visually responsive neurons with high firing rates, we provide a new picture of V1 that is substantially different from the textbook view.

The anatomical structure of V1 is characterized in terms of the canonical microcircuit.  This is essentially a structured, recurrent network, where each layer has specific inputs, projections, and feedback connections~\cite{Callaway1998, Mitzdorf1987}.  Single electrode recordings typically hone in on cells with vigorous visual responses that are evoked while an experimenter drifts bars or gratings of various types while advancing the electrode.  Experiments using multielectrode techniques having reduced sampling bias are increasingly common, yet models carried over from single cell recordings do not acknowledge the diversity in response properties.   While the field is moving rapidly towards the use of natural movie stimuli, these are often presented at much lower frame rates than the relevant time scale of the neural responses and may not probe the temporal dynamics of the cell~\cite{Butts2011}.  Impoverished stimuli have contributed to the notion of neurons being ``noisy'', or alternatively the view that the stimulus is only a small perturbation to ongoing recurrent dynamics.  As a consequence, model performance is often assessed on the basis of averaged peri-stimulus time histograms (PSTHs), giving the illusion of good model performance even when prediction on a single-trial basis is nearly impossible~\cite{Haslinger2012}.

These objections about biased stimuli and biased sampling~\cite{Olshausen2005} suggest a more exploratory approach to identify the true extent of our understanding of V1 function.  In this work, we attempt to address the two issues by modeling the responses of cells simultaneously recorded in all cortical layers in response to high temporal resolution natural movie stimuli.  We fit models to predict single trial responses to address the issue of trial-to-trial variability. To capture response properties that are not well described by simple or complex cell models, we use a nonlinear radial basis function classifier.  Our results show clear laminar differences in visual responsiveness and ease of modeling of neural responses.  Visual responsiveness is highest in cortical layers II/III and IV, where linear filter models perform best, suggesting a sampling bias for pyramidal cells in the superficial cortical layers in previous studies.

The rest of the paper is structured as follows: In Section 2 we describe the experimental methods, movie stimuli and the recording procedure, as well as preprocessing of the movies and spiking data for modeling. We then describe the models we fit: nonparametric (oracle) models, generalized linear models (GLMs) for simple and complex cell receptive fields and the radial basis function (RBF) classifier. This is followed by details on how models are evaluated using receiver operating characteristic (ROC) curves and correlation coefficient ($r^2$). 
In Section 3 we present results for fitting these models to data from 5 recording sessions, analyzing both single trials and PSTH data. We conclude with section 4.

% ----------------------------------
% METHODS
% ----------------------------------
\section{Methods}

\subsection{Recordings} 

%[STOLEN] from Pillow
Multi-electrode extracellular recordings spanning all cortical layers in anesthetized cat visual cortex were performed with 32-channel silicon polytrodes (Neuronexus). Multiunit activity was extracted using an adaptive thresholding algorithm and spike trains were binned in 20ms windows.

Custom natural scene movies were recorded at 300Hz frame rate. We focused on scenes with strong natural motion (e.g.\ animals moving in the frame) and avoided camera movement,  scene changes or ``cuts'' to reduce artifacts in the response such as evoked potentials due to sudden luminance changes. Long natural movies of 8-30 minute duration were presented in grayscale at $512 \times 384$ resolution and temporally downsampled to the CRT monitor refresh rate of 150Hz. A 30s segment of one of the movies was presented for 60 repetitions to compute a PSTH over repeated trials, the full movies were presented only once. 

\begin{figure}[t]
    \centering
    \includegraphics{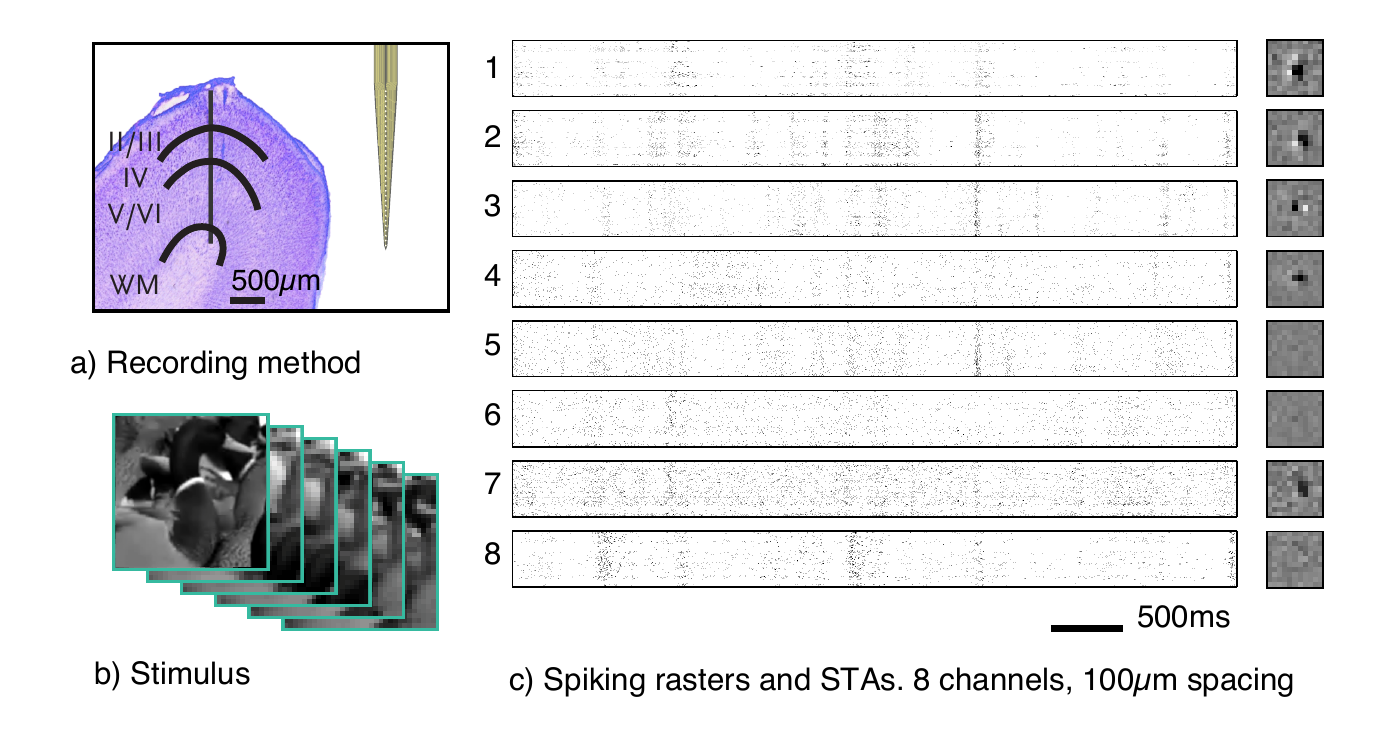}
    \caption{{\bf (a)} Polytrode recording method {\bf (b)} Example of cropped natural movie stimulus {\bf (c)} Spike triggered average in pixel and Fourier space and PSTH raster plot for two cells. Each STA panel corresponds to 20ms for 200ms total. The rasters are generated over 30s and 60 repetitions. The cells are simultaneously recorded from one microcolumn and responding to the same stimulus, yet there are strong qualitative differences in the linear receptive field and the spike rasters. The spike triggered averages in computed from a binary noise stimulus bear no clear relationship to the movie responses.}
    \label{fig:methods}
\end{figure}

In Fig.~\ref{fig:methods}a) we show the 32-channel recording device and an example recording location with the probe spanning the layers of a single cortical microcolumn. \ref{fig:methods}b) shows example frames of one of the natural movie stimuli used. \ref{fig:methods}c) gives an overview of the response properties of 8 simultaneously recorded units, spaced $100\mu m$ apart. The spiking rasters on the left span 5s and 60 repetitions of a natural movie stimulus. Spike triggered averages, on the right, are computed from a 20 minute binary noise movie presented at 30 fps that was not part of the following analysis. Closely spaced units reveal heterogeneous responses, for example unit 4 does not have the striated spiking raster of the neighboring cells, but a strong simple cell kernel. Units 3 and 4 have very similar spiking rasters, yet the receptive fields differ in polarity and size, while unit 5, which also shares a similar spiking raster, does not have a linear receptive field at all. %Our goal is to model this heterogeneity which is easily observed qualitatively, to better understand how .

% ----------------------------------
\subsection{Data Preprocessing}
We downsample the original $512 \times 384$ pixel movies to $32 \times 24$ pixels, and then extract a $10 \times 10$ pixel region around the center of the receptive field. Temporally, the movie was downsampled from 150 fps to 50 fps. 
For the Fourier power models (discussed below), power spectra were computed using an FFT of individual frames. 
A sequence of 10 subsequent frames, or 200ms, was then concatenated into a spatiotemporal stimulus vector and the dimensionality was reduced by PCA to 200 dimensions. Since there are strong temporal correlations between consecutive frames, this dimensionality reduction was in each case sufficient to retain above 90\% of the variance. During the PCA step, data were simultaneously whitened, and variable means subtracted. All models were trained on 13,000 samples and evaluated on another 8,000 samples from the same movie as well as a PSTH of 1500 samples computed over 60 repetitions. Cross validation was performed by composing the test and training set of randomly selected 30s blocks from the movie. The repeated movie segment was not part of the training movie. 

% ----------------------------------
\subsection{Models}

% ----------------------------------
\subsubsection{Nonparametric Model}
Before asking how to best model the visual responses, we need a way to determine what part of the response is due to visual stimulation and what activity is due to ongoing network activity or simply noise. To this end we use two types of nonparametric model. The first oracle model quantifies our uncertainty in computing explainable variance~\cite{David2005} from a small number of repetitions. We divide the 30s movie repetitions into 30 randomly selected repetitions that are averaged and used as the model, and the remaining 30 averaged repetitions as the data. The correlation coefficient between the two PSTHs represents the variance that can be explained by the visual stimulus. 

We use a second oracle model to quantify the visual response on a single trial basis. Here we compute a spiking probability from all but one of the trials and use the PSTH as a classifier to predict the response of the one hold-out trial using ROC curves (discussed below). While the explainable variance will always approach 100\% with enough trials, and thus essentially measures our uncertainty in computing a PSTH from a finite number of spikes, the area under the ROC curve measures how much a perfect model can capture about the response, and how much it is limited by trial-to-trial variability. 

\begin{SCfigure}
    \centering
    \includegraphics{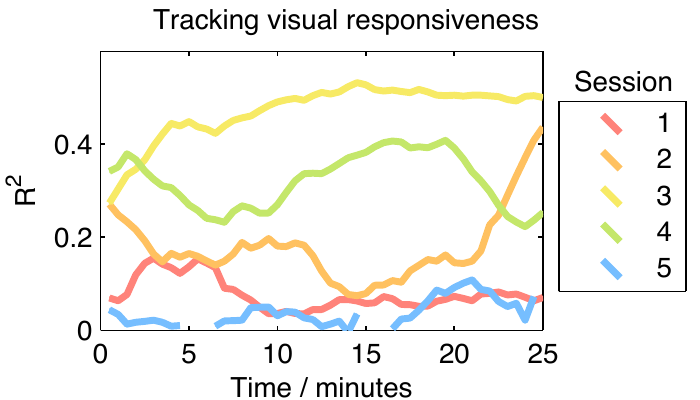}
    \caption{Using the oracle model on short 5 minute subsets, we can track changes of the cortical state. The plot shows average explained variance for all cells in a session. For sessions 1 and 5, the visual responses are weak throughout the repeated movie presentation. Other sessions, in particular 2, show marked fluctuations over the course of several minutes. }
    \label{fig:questionary}
\end{SCfigure}

These measure still depends on the stationarity of the recording, and as Fig. \ref{fig:questionary} shows, visual responsiveness can vary over a time scale of a few minutes. The squared variance is here computed on overlapping 5 minute windows (10 repetitions) over a 30 minute recoding of repeated 30s movies, and averaged over all cells in the session. As we will later see, in sessions where the responsiveness is  lower during the short movies repeats used to fit the oracle model than in the long movie from with the receptive field models are estimated, the RF models can actually outperform the oracle. 

%\comment{We could make a nice plot here showing how it drifts. Take only 20 trials and slide for 40 trials (20min of slide) divide into 10+10, (20 choose 10) gives a lot of permutations, and slide along by one trial. MAKE THIS PLOT AFTER FIRST PASS THROUGH PAPER!!}

% ----------------------------------
\subsubsection{Generalized Linear Model}
The LNP cascade is an example of a GLM, where we consider spike counts generated from a Poisson process with the firing rate $r(t)$ given by a linear filter followed by a point-wise nonlinearity:
\begin{align}
    r(t) &= f( \mb k^T \x(t)) \\
    \log p(D|X) &= \sum_t \left( n_t \log(r(t)) - r(t) dt \right)
\end{align}
where $\mb k$ is the linear filter, $x$ the stimulus vector and $f(.)$ a pointwise nonlinearity, which we fixed to an exponential. Equation 2 gives the log-likelihood of a Poisson process. From an optimization standpoint, the GLM shares the desirable property of the SVM that the estimation is a convex problem. Parameters can be estimated efficiently and convergence is guaranteed to be to a global rather than local optimum. 

In practice, the main limitation of the LNP is that the firing rate is determined by a linear filter response, and it is quite difficult to extend the GLM to a more nonlinear response without greatly increasing the dimensionality or losing the convexity of the model.
This shortcoming of the LNP model implies that it cannot capture nonlinearities such as the phase invariance properties of complex cells. To overcome this limitation, without changing the form of the model, we can perform nonlinear preprocessing on the stimulus terms and use this transformed stimulus as the input. Complex cells can be well modeled with the Fourier power spectrum of the stimulus~\cite{David2005}. 

Alternatively it is possible to perform a full expansion into a quadratic feature space, but this comes with a large increase in the stimulus dimensionality.
%~\cite{Gerwinn2010}. 
With these approaches, the convexity of the problem is conserved, which is not be the case if we introduced subunits with a squaring nonlinearity directly into the model~\cite{Vintch2012}. These two possibilities are not further explored here since the Fourier power model works well in practice and provides close to state of the art results in the neural prediction challenge\footnote{http://neuralprediction.berkeley.edu}, so the LNP model for simple cells and the Fourier power model for complex cells serve as the baseline against which we compare the nonlinear RBF models here. We also evaluated the Fourier phase-separated model~\cite{David2004}, but found it to perform no better than the FFT power model.
 We estimated the GLMs with an L$_1$ sparseness penalty on the elements of the linear filter, encouraging PCA components of the stimulus that do not contribute to the prediction to be set to zero. The optimal value of the sparseness parameter was determined by cross-validation for each cell individually.

% ----------------------------------
\subsubsection{RBF Kernel SVM}
Deviating from the classic RF estimation framework, we pose the neural response prediction as a binary classification problem. For each stimulus, we have two possible responses, ``spiking'' or ``not spiking''. This way we avoid making assumptions about a rate code. Since firing rates are low on average, it is rare that multiple spikes occur in one bin. 
The Support Vector Machine (SVM) 
%~\cite{Smola1998}
is a linear classifier that constructs a maximum margin hyperplane from a subset of data points. This is an effective way to regularize with only few training examples, making it well suited for estimation in high-dimensional, nonlinear spaces. The RBF kernel was used to estimated a nonlinear classifier. Intuitively this can be linked to clustering with a mixture of Gaussians with means centered on data points, and spherical covariances. This allows representation of highly nonlinear responses, as long as they trace out a smooth manifold in the kernel space. In practice the main drawback of the method is that complexity scales quickly with the size of input data, so estimation is not feasible for very large date sets. The SVM estimation was performed using the LibSVM package~\cite{Chang2011}.

% ----------------------------------
\subsection{Model Evaluation}

We are primarily interested in the single trial prediction performance of the models. Since we can upper bound it with the oracle model this method is more informative than explained variance, which ignores trial-to-trial variability, or likelihood, which is hard to obtain a practical upper bound on. 
We obtain single trial predictions from the probabilistic regression models by thresholding the outputs in the same way as the binary classfiers.
To compute explained variance, we can also cast the SVM in a probabilistic framework and measure correlation with the PSTH. We use Platt's histogram method~\cite{Platt1999} to obtain probabilities from the decision function of the SVMs. This method computes histograms of the decision function of the SVM for the positive and negative class, and fits a sigmoid function to the ratio of positive to all data points for a given bin. %This probability is not as elegant as a maximum likelihood estimate, but it can be seen intuitively that by capturing the frequency of misclassifications for a given value of the decision function, it quantifies the confidence of the classifier. 

%\subsubsection{ROC curves}
The ROC curve of a classifier is obtained by varying the threshold to trace out a set of models. At the lowest threshold, we classify all data points as positive (spiking), and at the highest, as all negative (non-spiking). Plotting true positives against false positives as we trace out this set, a perfect model will result in an area under the curve (AUC) of one, as all positives will always be true positives. For a model that performs at chance level, the AUC will be 0.5 as the frequency of positive hits will just follow the fraction of positives in the data. We traced out the curves by determining the two extreme values of the threshold that would result in all positives or negatives and then numerically integrate over the curve in 10,000 steps. For more details of the method, see~\cite{Truccolo2009}, where ROC curve analysis has been performed in the context of spike prediction from a population of cells. 

%\subsubsection{R-squared}
To evaluate the performance of probabilistic models, a common method is the percentage of explained variance used by Gallant et at.~\cite{David2004}. This is not a meaningful metric for single trial prediction with binary variables since it assumes Gaussian noise, but the correlation coefficient $r^2$ is useful as a metric for the distance between model and a PSTH over repeated trials. 
While the the area under the ROC curve is bounded by the stochastic component of the response such as noise or ongoing neural dynamics, for the variance explained to be one it is sufficient for the model to captures only the visually evoked part of the response. Values of $r^2$ reported in the literature are typically below 50\% for recordings in response to natural movies~\cite{David2005}, but are strongly dependent on bin size and number of trials.

% ----------------------------------
% Results
% ----------------------------------
%\clearpage
\section{Results}

% ROC curves
\begin{figure}
    \centering
    \includegraphics{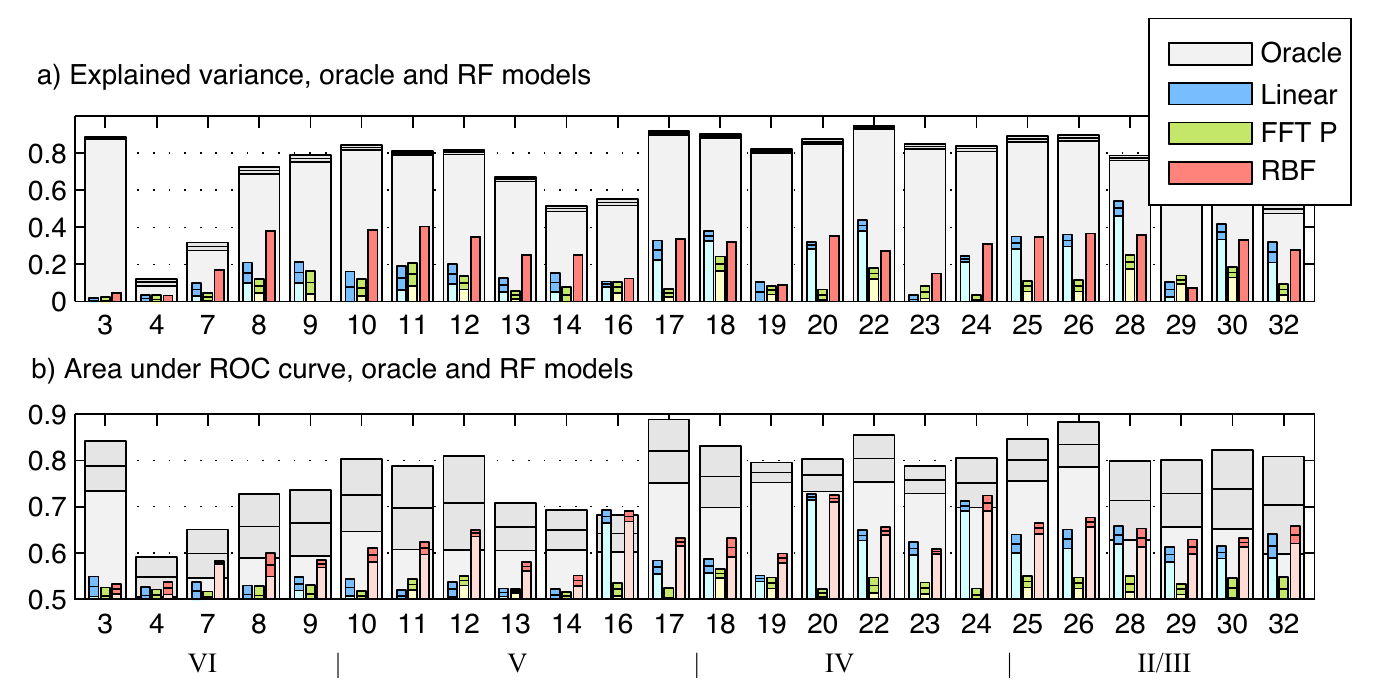}
    \caption{Comparison of RBF kernel SVM, linear GLM and Fourier power GLM for one of the five recording sessions. In {\bf(a)} we compare explained variance of the PSTH, in {\bf(b)} area under the ROC curve. In both cases, an oracle model is computed from the short movie PSTH. 
The RF model generally fall far short of the oracle performance, even discounting trial-to-trial variability the majority of the visual response is not captured. The RBF model is the best of the RF models, followed by the linear pixel-space GLM. Error bars on all models indicate standard deviation computed using cross-validation.    }
    \label{fig:bar}
\end{figure}

% NEED TO REWRITE THE FIRST PARAGRAPH WITH THE HETEROGENEITY RESUTS. THEN STICK MORE OR LESS TO THE SCRIPT

We report on $N=103$ cells from 5 recording sessions, selected to have above 1Hz average firing rate during presentation of the long movie. RBF kernel, pixel GLM and Fourier power GLM models were estimated independently for each of the cells. 
We estimated models on a subset of 13,000 samples with a holdout set of of 8,000 samples randomly selected for 10-fold cross-validation. Hyperparameters of  the SVM (regularization parameter $C$ and RBF kernel scale parameter $\gamma$) were estimated for each cell by grid-search on a logarithmic scale with a factor of 10 spacing. The GLMs were estimated similarly, with a search for optimal sparseness penalty $\lambda$, bracketing it to a factor of 2. We used the optimal parameters for both ROC and $R^2$ comparisons, even if the optima were reach with different hyperparameters.  

The cells included in the study had a wide range of visual responsiveness, quantified by the AUC of the oracle model ROC curves. Out of the visually responsive cells, there are strong difference in how well these responses can be captured by the different receptive field models we estimated. An example for one recording session is shown in Fig.~\ref{fig:bar}, with simultaneously recorded cells ordered from deep to superficial. In  Fig.~\ref{fig:bar}a) we show the explained variance for each of the cells, with the RF models plotted in front of wide bars corresponding to the oracle model. 
This ``explainable variance'' quantifies uncertainty in estimating the PSTH from a finite number of trials. For cells with low spike counts during the repeated movie presentation, or weak visual responses, the oracle performance drops close to zero. While the RBF model never exceeds 30\% explained variance in this particular recording, it does perform as well as the oracle for the most superficial cells. Note that it is possible for the RF model to outperform the oracle if the long movie for which the RF model is estimated was presented during a more responsive cortical state than the repeated movies from which the PSTH is estimated. 
The single trial prediction in Fig.~\ref{fig:bar}b) follows the same trend as the explained variance, but the variation in the single trial oracle is increased reflecting trial-to-trial variability. We now see in deeper layers and particular layer IV that RF model performance approaches the oracle. The models are trained only on single trial data, so relative to oracle performance they perform better in this prediction task. 
There is a clearly visible layer dependence in the performance of all models. The most visually responsive units are localized in layers II/III and IV, and relative to the oracle these responses are also better captured by RF models. Cells in the deep cortical layers are less responsive and harder to model.

\begin{figure}
    \centering
    \includegraphics{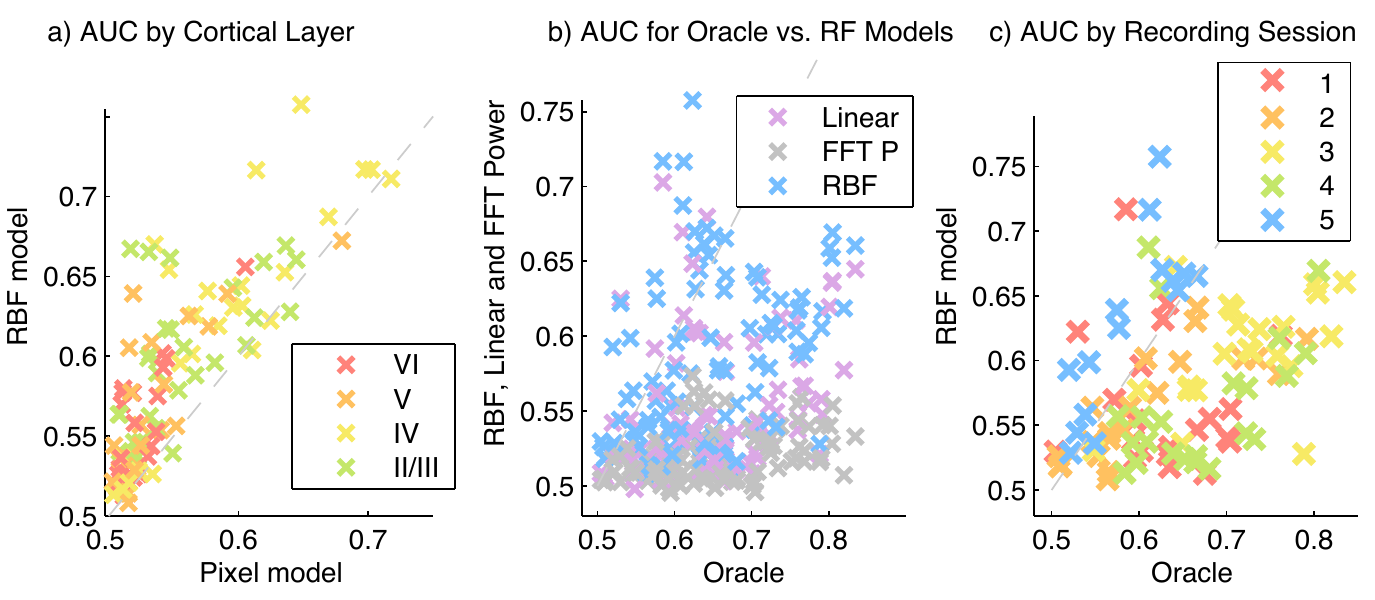}
    \caption{{\bf(a)} Scatter plot showing model performance in the different cortical layers. The RBF kernel outperforms the liner pixel model. Cells in superficial layers are better predicted than in deeper cortical layers.   {\bf(b)} Comparison of the different RF models with the oracle. For most cells, the RF models perform only slightly above chance (AUC=0.5). While the oracle model does better, performance is highly variable showing that many cells do not have a strong visually evoked response. {\bf(c)} shows a subset of (b), the RBF model, but color coded by recording session, showing the global differences in responsiveness.  }
    \label{fig:scatter}
\end{figure}

The scatter plot in Fig.~\ref{fig:scatter} summarizes these results for all 106 cells. Fig.~\ref{fig:scatter}a) compares the ROC performance of the RBF kernel to the linear filter model, showing that the nonlinear model outperforms the linear one on 94/103 cells. Still the overwhelming majority of cells are not well modeled with the AUC staying close to the chance level of 0.5. Setting an arbitrary threhold of $AUC=0.6$, we find that 44\% of layer II/III cells, 62\% of layer IV, 29\% of layer V and only 8\% of layer VI cells are visually responsive, which agrees well with the anatomical role of layer IV as the input layer that would be expected to have the most linear responses. 
In Fig.~\ref{fig:scatter}b), we compare the linear, Fourier power and RBF models to the oracle for all cells. The correlation between oracle and RF model is low, which shows the surprisingly strong effect cortical state changes have on the model performance. As we showed in \ref{fig:questionary}, the visual responsiveness changes dramatically over the time course of minutes. With approximately one hour elapsing between the presentation of the long movie from which the RF models are estimated, and the repeated movie that the oracle is computed from, it is not surprising that in some cases the RF model actually offers better prediction than the oracle. On average, though, the oracle provides captures significant visual responses while the RF models are slightly closer to chance. 
Comparing the different RF models we see that the RBF and linear models are close in performance and much better than the Fourier power model. Units where the FFT outperformed the pixel model were usually much better captured by the RBF model, suggesting that phase invariance is not the dominant nonlinearity in these cells. 
\ref{fig:scatter}c) is a subset of b) showing only the RBF responses, color coded by recording session. This lets use evaluate responsiveness of the session during the short movie presentation (oracle) and the long movie (RBF model). While there are clear differences in the overall responsiveness, all sessions include non-responsive cells and cells that can only be captured by the oracle, but not the RBF models.

%The effect of spike count is summarized in Fig.~\ref{fig:scatter}c) which directly compares the oracle performance against spike count. Note that we select the training set for the RF models based on a minimum spike count, but there is no such criterion for the PSTH, thus sessions 4 and 8 have very low spike counts. This explains the cells from session 1 where the oracle performs worse than than the RF model.  

\begin{figure}
    \centering
    \includegraphics{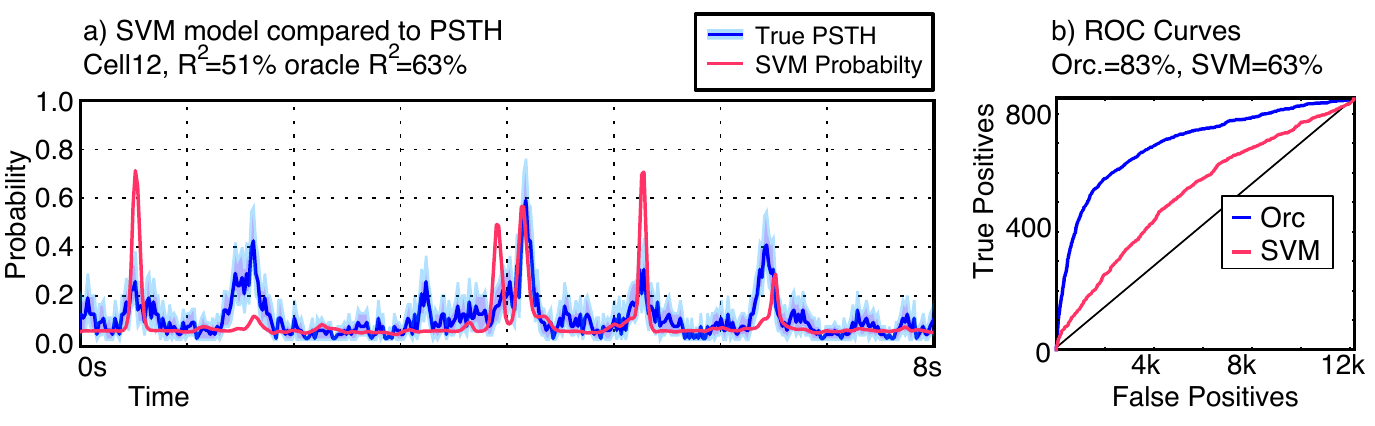}
    \caption{{\bf(a)} Comparison of probabilistic model output with the PSTH collected over 60 repeated trials over 30s of data. The shaded region around the PSTH indicates 5\% confidence intervals computed by bootstrapping. Even though a correlation coefficient of 51\% would generally be considered high (even more so as the Oracle model only explains 63\% variance) it is clear that many of the sharp temporal transients are not captured by the model.
    {\bf(b)} ROC curve analysis to characterize model fit. Varying the threshold of the classifier, we trace out a curve varying the number of true and false positives in the data. In contrast to percentage error, this method is not sensitive to the total frequency of positive vs.\ negative examples. An AUC=1 could only be reached if there was no trial-to-trial variability.  }
    \label{fig:psth_roc}
\end{figure}

A more detailed example for a single unit is shown in Fig.~\ref{fig:psth_roc}, with a) the PSTH trace (blue, with 5\% and 95\% confidence intervals estimated by bootstrapping in light blue) and the best SVM prediction for this cell. Although the correlation in this case is high at 51\% and very close to oracle performance at 63\%, visually the fit is not as good as most of the sharp transient temporal structure, that arguably represents the most relevant part of the response, is not captured well. For the same cell, Fig.~\ref{fig:psth_roc}b) shows the ROC curves for RBF and oracle. Taking into account the actual count values for the positives, it becomes apparent that even a seemingly high AUC above 80\% will result in more false positives than true positives at the knee of the curve, so even the oracle model captures only little of the single trial response. 
 
% ----------------------------------
% Conclusion
% ----------------------------------
\section{Discussion}

% part one: recording bias
% how to structure this:
% - Suprising number of not responsive cells
% - Surprising because not in the literature
% - not in the literature because it's normally hard to corectly attribute
% - therefore to be save things are attributed to state changes 
% - and therefore the bias is created.
While the true test for our models would require recordings from the awake, behaving brain under natural viewing conditions, we have gone part of the way by attempting to use a visual stimulus that is as natural as possible.
Bearing in mind that the vast majority of our knowledge about V1 is based on recording from anesthetized preparations using simplified stimuli, the results described here still provide a useful basis for comparison. 
 eexamining the anesthetized condition, but explicitly avoiding to ``cherry pick'' cells with the silicon polytrode recording technique, we expose the full gamut of neurons and show that cells behaving in the way the standard model predicts, while present, are in the minority. 

%By using silicon polytrodes, we serendipitously recorded from a heterogeneous population of cells with a surprisingly large variability in responses. 
We offer the following explanation why these cells may not be recorded or analyzed in a single electrode experiment: When recording a neuron that is not visually responsive in isolation, it can not be distinguished if the (lack of) response is due to cell type, cortical state, or both.
%Cortical state changes are unavoidable in anesthetized preparations. This can potentially lead to confusion 
This confusion between non-classical cell types and cells in a temporarily unresponsive state, while unavoidable in anesthetized preparations, thus leading to a bias to record or analyze cells that pass sanity checks for responsive states.
Here the simultaneous recording helps us to disentangle these alternatives, since we can ensure a visually responsive cortex as long as some fraction of the cells are responsive. We find cells that are not responsive in all recording sessions.
Cells that are not visually responsive or well-modeled by simple and complex cell models thus appear to be an important factor influencing recordings alongside cortical state changes or recording instabilities.
In particular, systematic differences in the responses of different cortical layers suggest that each layer has properties that need to be analyzed and modeled separately.
A focus on atypical cells, in particular in the deep cortical layers, will be crucial for a better and more complete understanding of primary visual cortex.

% part two: natural movie stimuli
The use of natural movie stimuli, which are not designed to evoke reliable responses like grating stimuli, further leads to recordings that do not lend themselves to characterization with existing RF mapping methods. Firstly, many state of the art methods make strong assumptions about known properties, e.g. for simple cells~\cite{Park2011}. In addition, most work on modeling the stimulus response of V1 cells has focused on experiments using white noise stimulation which leads to qualitatively different responses to natural stimuli~\cite{David2004}, as many of the complex surround effects are not present~\cite{Haslinger2012}. Natural movie stimuli make it less straightforward to apply complex cell models based on the Energy model~\cite{Adelson1985}, such as spike triggered covariance~\cite{Touryan2002} or the subunit model of~\cite{Vintch2012} to our data. The drawback of the Fourier power model we use is that it is completely invariant to position and therefore depends on manually determining the boundaries of the receptive field for best performance, however we found no improvement by fine-tuning the size of the window in which the receptive field was computed.  

A weakness of our models is that they do not include surround effects like gain control, which play a large role in V1 under artificial~\cite{Heeger1996} and natural~\cite{Coen-Cagli2012} viewing conditions. It is difficult to model these effects in natural movies since including the surround would drastically increase the already high stimulus dimensionality. A treatment of gain control and surround inhibition effects could also be based on a population-level model of activity~\cite{Pillow2008}, which would prove worthwhile with the dataset presented here.

% conclude
%In conclusion, we have characterized the visual response properties of an unbiased population of cells from a cortical column of V1 using natural movies carefully designed to mimic natural visions as closely as possible in their spatiotemporal statistics. Presentation at 150 fps allows us to to analyze temporal structure on the 10ms time scale. 

\bibliographystyle{unsrt} % plain.bst or apalike
\bibliography{/Users/urs/Dropbox/mendeley} % mendeley 

\end{document}